\documentclass{moriond}

\usepackage{amsmath}
\usepackage{amssymb}
\usepackage{graphicx}
\usepackage{xspace}
\usepackage{color}
\usepackage{slashed}
\usepackage{units}
\usepackage[dvipsnames]{xcolor}

\usepackage{hyperref}

\newcommand{\gambit}{\textsf{GAMBIT}\xspace}

\newcommand{\colliderbit}{\textsf{ColliderBit}\xspace}

\newcommand{\specbit}{\textsf{SpecBit}\xspace}
\newcommand{\decaybit}{\textsf{DecayBit}\xspace}

\newcommand{\scannerbit}{\textsf{ScannerBit}\xspace}

\newcommand{\GB}{\gambit}

\usepackage{amssymb}

\def\be{\begin{equation}}
\def\ee{\end{equation}}
\def\bea{\begin{eqnarray}}
\def\eea{\end{eqnarray}}

\newcommand{\Photo}{}

\begin{document}
\vspace*{4cm}
\title{EXPLORING LIGHT SUPERSYMMETRY WITH GAMBIT}

\author{ A.\ KVELLESTAD, on behalf of the GAMBIT Collaboration }

\address{Department of Physics, Imperial College London, Blackett Laboratory,\\ Prince Consort Road, London SW7 2AZ, United Kingdom}

\maketitle\abstracts{
I summarize a recent study by the GAMBIT Collaboration in which we investigated the combined collider constraints on the chargino and neutralino sector of the Minimal Supersymmetric Standard Model. Through a large fit using \GB we found that current ATLAS and CMS results with 36\,fb$^{-1}$ of 13\,TeV LHC collision data do not provide a general constraint on the lightest neutralino and chargino masses. Further, we found that a pattern of excesses in some of the LHC analyses can be fit in a subset of the model parameter space. The excess has an estimated local significance of $3.3\sigma$ based on the 13\,TeV results alone, and 2.9$\sigma$ when 13\,TeV and 8\,TeV results are combined.
}

Due to its potential for solving problems such as the hierarchy problem and dark matter, weak-scale supersymmetry (SUSY) has long been among the best motivated scenarios for experimentally accessible physics beyond the Standard Model (SM). A commonly studied realisation of SUSY is the Minimal Supersymmetric Standard Model (MSSM), in which large parametric freedom is used to express ignorance about what mechanism is responsible for breaking SUSY. As a consequence the MSSM parameter space encompasses a wide range phenomenological scenarios, making it both a useful and challenging target for experiments.

Searches for SUSY particles by the ATLAS and CMS experiments at the LHC are typically optimised on so-called ``simplified models''. In these models only the two or three lightest SUSY particles are assumed to contribute to the targeted signal. This assumption reduces the size of the relevant parameter space to a manageable size for experimental optimisation---typically two or three masses and/or branching ratios. A downside of this approach, however, is that it is not clear to what extent the resulting experimental constraints on the simplified models also apply in more realistic SUSY realisations such as the MSSM. This question motivated a recent study by the GAMBIT Collaboration, in which we investigated the constraints on the full neutralino and chargino sector (the \textit{electroweakino} sector) of the MSSM resulting from a combination of 13\,TeV LHC results and other collider constraints.\cite{EWMSSM}

For this study we assumed that all sparticles except the neutralinos and charginos are too heavy to impact current LHC searches. This assumption limits the relevant electroweak MSSM parameter space to only four free parameters: the bino mass parameter $M_1$, the wino mass parameter $M_2$, the Higgsino mass parameter $\mu$, and the ratio of the vevs for the two Higgs doublets $\tan\beta$. We name this four-parameter model EWMSSM.

Using \GB,\cite{gambit} and in particular the \specbit, \decaybit and \colliderbit modules,\cite{SDPBit,colliderbit} we performed a fit of the EWMSSM to recent results from 13\,TeV ATLAS and CMS searches for charginos and neutralinos. See Ref.~1 for details of the included analyses. 
At each sampled EWMSSM parameter point we ran full Monte Carlo simulations of the LHC searches and used the predicted signal rates to formulate a proper joint likelihood for our fit. In this likelihood function we also included other relevant collider observables, namely the Z and Higgs invisible decay widths and SUSY cross-section limits from LEP. To efficiently explore the parameter space we used the \textsf{Diver} differential evolution sampler via \scannerbit.\cite{scannerbit} After the parameter scan we repeated our LHC simulations with higher event statistics for all parameter samples within the identified best-fit parameter regions, going up to 64 million events per point for the 500 highest-likelihood points.

\begin{figure*}
  \centering
  \includegraphics[height=0.4\columnwidth]{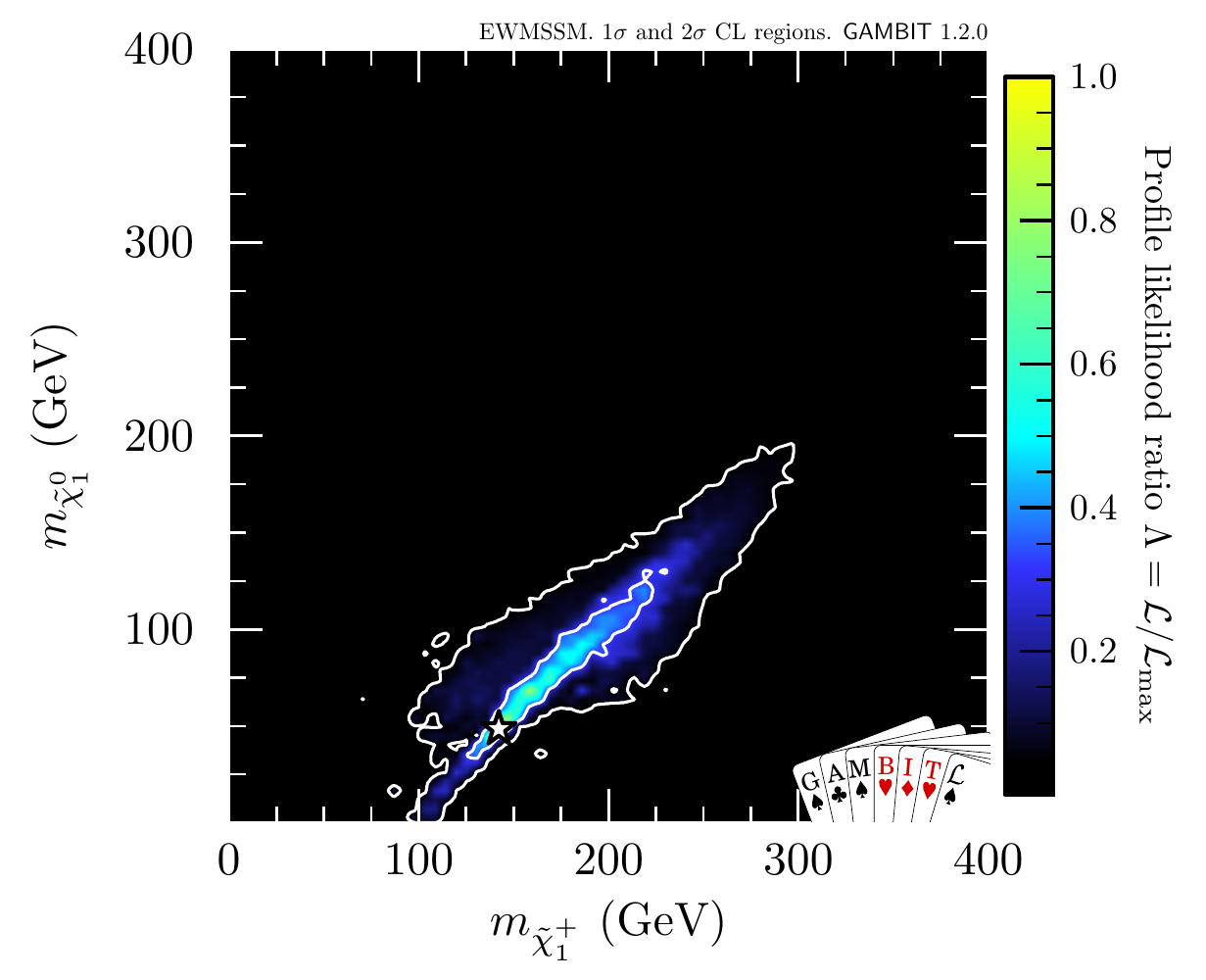}
  \includegraphics[height=0.4\columnwidth]{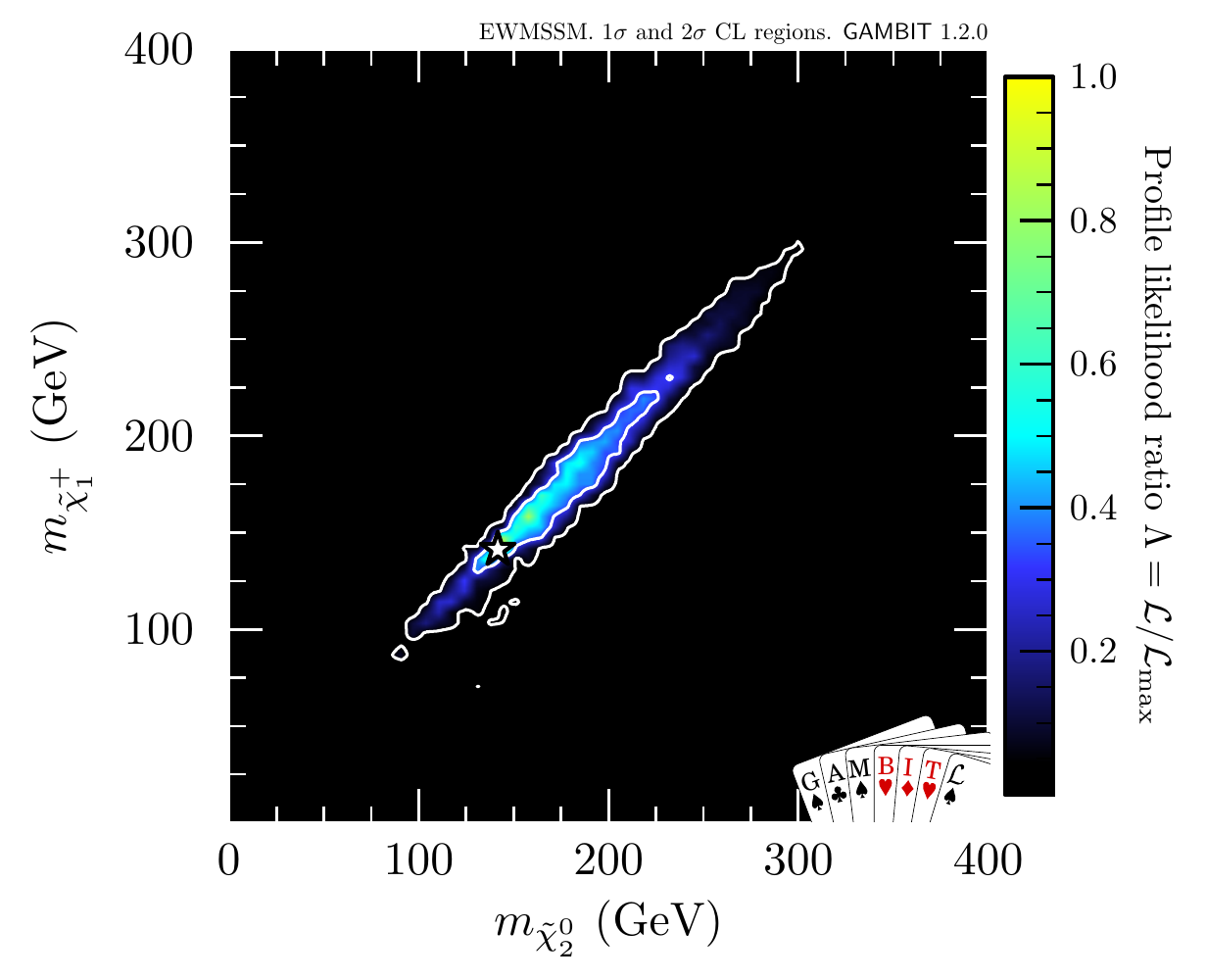}\\
  \includegraphics[height=0.4\columnwidth]{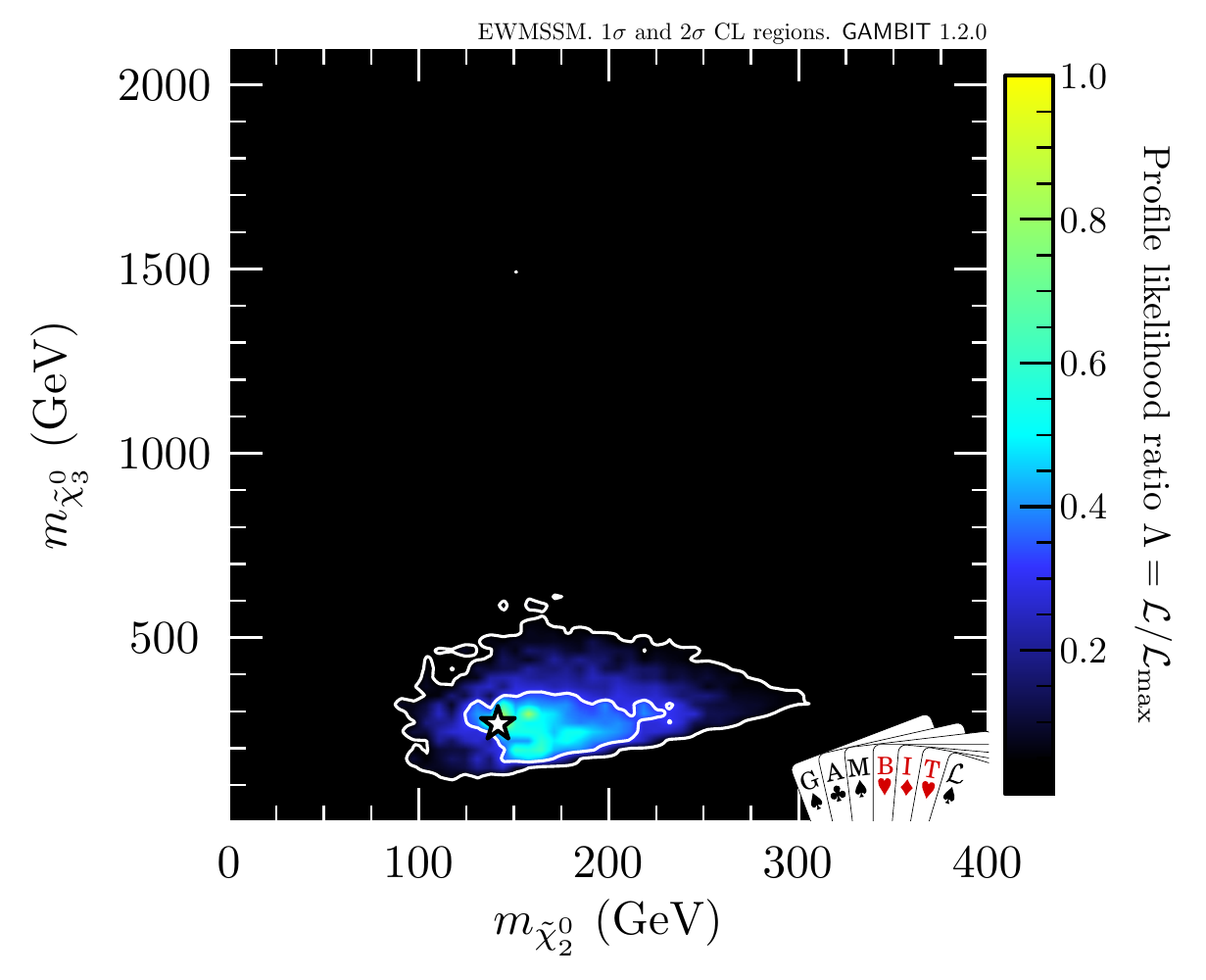}
  \includegraphics[height=0.4\columnwidth]{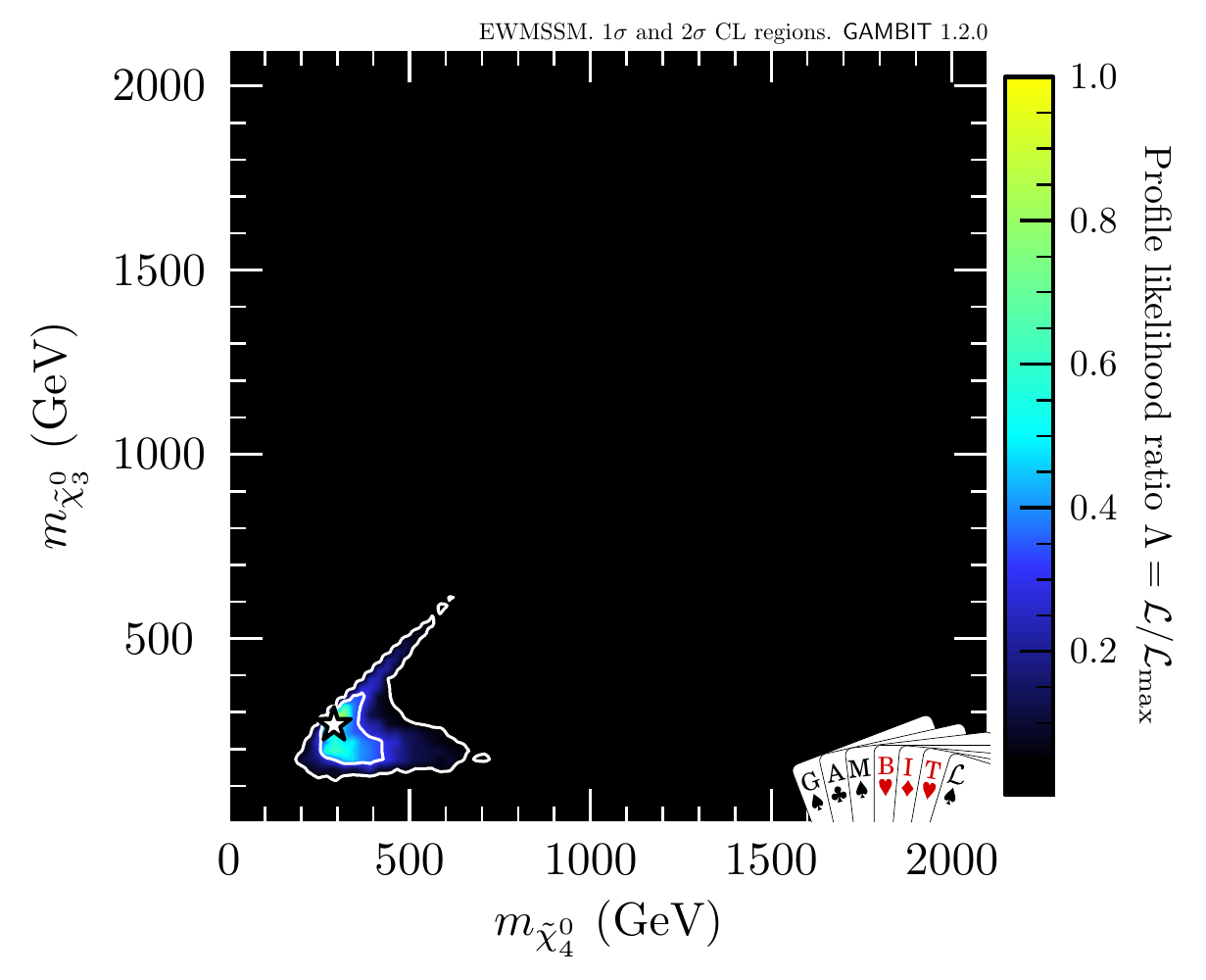}
  \caption{
  The profile likelihood ratio shown for the following planes of neutralino and chargino masses: $(m_{\tilde{\chi}_1^{\pm}},m_{\tilde{\chi}_1^0})$, $(m_{\tilde{\chi}_2^0},m_{\tilde{\chi}_1^{\pm}})$, $(m_{\tilde{\chi}_2^0},m_{\tilde{\chi}_3^0})$ and $(m_{\tilde{\chi}_4^0},m_{\tilde{\chi}_3^{0}})$. The white contours outline the $1\sigma$ and $2\sigma$ confidence regions and the best-fit point is indicated by the white star.
  }
  \label{fig:pole_masses_2D}
\end{figure*}

The main results of our analysis are displayed in Fig.~\ref{fig:pole_masses_2D}, which shows the profile likelihood ratio in different planes of neutralino and chargino masses. The star indicates the best-fit point while the white contours outline the 1$\sigma$ and 2$\sigma$ confidence regions. We see that, when combined, the current collider data prefers a specific pattern of low-mass solutions in the EWMSSM, with the lightest neutralino below $\sim200$\,GeV (top left panel) and the heaviest neutralino below $\sim 700$\,GeV (bottom right panel) at the 2$\sigma$ level. In these scenarios the lightest neutralino is dominantly bino, but with a non-negligible Higgisino or wino component. Another characteristic of the best-fit region is the presence of two $\gtrsim m_Z$ gaps in the electroweakino mass spectrum.

There are a number of small data excesses in some of the LHC searches, 
most importantly the ATLAS SUSY searches for 2-, 3- and 4-lepton final states.\cite{Aaboud:2018jiw,Aaboud:2018sua,Aaboud:2018zeb} We found that these excesses can be simultaneously fit in the EWMSSM, explaining the preference for low-mass solutions in our result. 
The excesses are seen in signal regions that target leptons coming from decays of on-shell SM gauge bosons. This is why our fit favours spectra with mass differences large enough to produce on-shell Z's and W's in the sparticle decays. The LHC search displaying the strongest tension with our best-fit scenario is the CMS search for multi-lepton final states,\cite{CMS-PAS-SUS-16-039} as further discussed in Ref.~1. Through dedicated Monte Carlo simulations of our likelihood ratio test statistic, we estimated the local significance of the combined excess to be 3.3$\sigma$ when using our best-fit EWMSSM point as the signal hypothesis. 

Our original fit did not include simulations of 8\,TeV SUSY searches, as this would essentially have doubled the computational expense of our study. To investigate the impact of 8\,TeV results on the EWMSSM scenarios preferred by our fit we therefore post-processed the parameter samples in the 1$\sigma$ region with simulations of several 8\,TeV ATLAS and CMS analyses. The result was a small shift of the best-fit point towards higher masses---the best-fit value for the lightest neutralino mass moved from 49.4\,GeV to 67.3\,GeV--- accompanied by a reduction in the local significance, from 3.3$\sigma$ to 2.9$\sigma$.

Regardless of whether the current excess turns out to be an early hint of new physics or not, our result highlights a caveat regarding the interpretation of constraints derived for simplified SUSY models: A mass hypothesis for the lighter electroweakinos that is excluded in a simplified model, may still be perfectly viable---and even preferred---within the more general MSSM electroweakino sector. There are several reasons for this. First, the additional signal processes introduced by the heavier neutralinos and charginos can provide improved fits to other analyses/signal regions, which in a proper statistical combination can compensate for the bad fit to the analysis/signal region that excluded the simplified model scenario. Second, for LHC analyses that only make use of the signal region with the best expected exclusion sensitivity, the heavier electroweakino states can affect which signal region is identified as the most sensitive one. Third, even though a simplified model scenario and an MSSM scenario may have identical masses for the lighter electroweakinos, the bino/wino/Higgsino admixture of these states may differ, affecting the production cross-sections and branching ratios. All these effects are at play in our fit.

\section*{Acknowledgments}

I thank my colleagues in the GAMBIT community for collaboration on the work presented here, and acknowledge PRACE for granting GAMBIT access to the Marconi supercomputer at CINECA, Italy.

\section*{References}

\bibliography{refs}

\end{document}